\documentclass[11pt,aps,prd,showpacs,amsmath,amssymb,floatfix,nofootinbib,superscriptaddress]{revtex4-2}

\usepackage{graphicx}  
\usepackage{multirow}
\usepackage[dvipsnames]{xcolor}
\usepackage[colorlinks=true,allcolors=BlueViolet]{hyperref}
\usepackage{multirow,bigdelim}
\usepackage[dvipsnames]{xcolor}

\usepackage{latexsym}
\usepackage{amsfonts}
\usepackage{amsmath,amssymb}
\usepackage{slashed}
\usepackage{color}
\usepackage{orcidlink}
\usepackage{supertabular} 
\usepackage{epsfig}
\usepackage{lipsum}

\usepackage[normalem]{ulem}

\newcommand{\olsi}[1]{\,\overline{\!{#1}}}

\newcommand{\bqn}{{\boldsymbol q}_n}
\newcommand{\bq}{\boldsymbol q}

\newcommand{\OPE}{\text{OPE}}

\newcommand{\uestc}{\affiliation{School of Physics, University of Electronic Science and Technology of China, Chengdu 611731, China}}

\newcommand{\ific}{\affiliation{Instituto de F\'isica Corpuscular (centro mixto CSIC-UV), \\
Institutos de Investigaci\'on de Paterna, Apartado 22085, 46071, Valencia, Spain}}

\newcommand{\itp}{\affiliation{Institute of Theoretical Physics, Chinese Academy of Sciences, Beijing 100190, China}}

\newcommand{\ucas}{\affiliation{School of Physical Sciences, University of Chinese Academy of Sciences, Beijing 100049, China}}

\newcommand{\scnt}{\affiliation{Southern Center for Nuclear-Science Theory (SCNT), Institute of Modern Physics,\\ Chinese Academy of Sciences, Huizhou 516000, China}}

\begin{document}

\title{\boldmath Predicting $B^{(*)}\olsi D{}^{(*)}$ states from a reanalysis of lattice QCD data}

\author{Pan-Pan Shi\orcidlink{0000-0003-2057-9884}} \email{Panpan.Shi@ific.uv.es}
\ific

\author{Miguel Albaladejo\orcidlink{0000-0001-7340-9235}}\email{Miguel.Albaladejo@ific.uv.es}
\ific 

\author{Meng-Lin~Du\orcidlink{0000-0002-7504-3107}}\email{du.ml@uestc.edu.cn} \uestc \scnt

\author{Feng-Kun~Guo\orcidlink{0000-0002-2919-2064}}
\email{fkguo@itp.ac.cn}
\itp\ucas\scnt

\author{Victor Montesinos\orcidlink{0000-0002-6186-2777}}\email{victor@itp.ac.cn}\itp

\author{Juan Nieves\orcidlink{0000-0002-2518-4606}} \email{jmnieves@ific.uv.es}
\ific


\begin{abstract}
We reanalyze the lattice QCD data for the isoscalar $\bar b\bar c ud$ systems with $J^P=0^+$ and $1^+$ at $m_{\pi}\simeq 220$ MeV, incorporating both short- and long-range interactions constrained by heavy-quark spin and chiral symmetries. Using the interaction fitted to the lattice results, we predict a spectrum of isoscalar $B^{(*)}\olsi D{}^{(*)}$ molecular states with quantum numbers $J^P=0^+$, $1^+$, and $2^+$. The predicted $J^P=0^+$ and $1^+$ $B\olsi D$ and $B^*\olsi D$ bound states are consistent with the lattice results. Furthermore, the two resonances reported by the lattice collaboration in the $J^P=0^+$ and $1^+$ $\bar b\bar c ud$ channels can be naturally interpreted as $B^*\olsi D{}^*$ and $B\olsi D{}^*$ hadronic molecules, respectively. We also predict two previously unreported $B^*\olsi D{}^*$ states with $J^P=1^+$ and $2^+$: a bound state and a shallow state whose pole may correspond to either a virtual state or a bound state, respectively. Our results provide a unified interpretation of the lattice spectrum and highlight the interplay between short- and long-range interactions in the dynamics of the $B^{(*)}\olsi D{}^{(*)}$ systems.
\end{abstract}

\maketitle

\section{Introduction}

The past two decades have witnessed the discovery of numerous hadrons that lie beyond the conventional quark model, in which mesons and baryons are described as quark-antiquark and three-quark states, respectively. Considerable experimental and theoretical efforts have been devoted to uncovering the nature of these unconventional hadrons, as reviewed in Refs.~\cite{Hosaka:2016pey,Esposito:2016noz,Lebed:2016hpi,Ali:2017jda,Olsen:2017bmm,Guo:2017jvc,Liu:2019zoy,Guo:2019twa,Brambilla:2019esw,Chen:2022asf,ParticleDataGroup:2024cfk,Liu:2024uxn,Chen:2024eaq,Wang:2025sic,Dai:2026fkg}. Among them, hadrons with exotic flavor content and quantum numbers are particularly compelling, as they cannot mix with conventional mesons or baryons in Fock space and therefore constitute unambiguous candidates for exotic multiquark states.

One of the most promising candidates for an exotic hadron is the doubly charmed tetraquark $T_{cc}(3875)^+$. It was discovered by the LHCb Collaboration in the $D^0D^0\pi^+$ invariant mass spectrum with quantum numbers $I(J^P)=0(1^+)$~\cite{LHCb:2021auc,LHCb:2021vvq}. Its quark content, $cc\bar u\bar d$, unambiguously establishes its multiquark nature, while its proximity to the $D^{*+}D^0$ threshold and decay properties strongly support its interpretation as a hadronic molecule dominated by the $D^{*+}D^0$ component~\cite{LHCb:2021auc,Feijoo:2021ppq,Albaladejo:2021vln,Du:2021zzh,Albaladejo:2022sux}.

Motivated by the discovery of the $T_{cc}^+$, its analogues with quark content $\bar b\bar c u d$ or $bc\bar u\bar d$ have attracted considerable theoretical attention. These systems have been investigated within a variety of theoretical frameworks. Quark-model calculations predict tetraquark masses either above~\cite{Silvestre-Brac:1993zem,Ebert:2007rn,Eichten:2017ffp,Park:2018wjk,Lu:2020rog,Deng:2018kly,Braaten:2020nwp,Song:2023izj,
Weng:2021hje} or below~\cite{Lee:2009rt,Karliner:2017qjm,Carames:2018tpe,Deng:2018kly,Yang:2019itm,Tan:2020ldi,Guo:2021yws,Weng:2021hje} the relevant $B^{(*)}\olsi D{}^{(*)}$ thresholds, while QCD sum-rule analyses generally place their masses close to threshold~\cite{Chen:2013aba,Agaev:2018khe}. In the hadronic molecular picture, meson-exchange models predict isoscalar $B^{(*)}\olsi D{}^{(*)}$ bound states with quantum numbers $J^P=0^+$, $1^+$, and $2^+$ and binding energies ranging from a few to several tens of MeV~\cite{Li:2012ss,Sakai:2017avl,Ren:2026wzt}. These states have also been investigated within a framework based on heavy-quark spin symmetry (HQSS), which predicts shallow isoscalar and isovector molecular states~\cite{Abreu:2022sra}. More recently, the $S$-wave $\olsi B D$ molecular state has been proposed as a potential target for searches through its weak decay, by reconstructing the $\pi^+K^-\olsi B{}^0$ invariant-mass distribution~\cite{Liu:2024lmv}.

Several lattice QCD (LQCD) studies have investigated the $\bar b\bar c u d$/$bc\bar u\bar d$ systems, revealing a rather diverse picture of their low-energy spectrum. The first calculation, using both tetraquark and meson-meson interpolating operators, reported a ground state in the isoscalar $J^P=1^+$ channel lying $50(11)$ and $22(7)$ MeV below the $B^*\olsi D$ threshold at pion masses $m_\pi\simeq164$ and $299$ MeV, respectively, providing evidence for an exotic state~\cite{Francis:2018jyb}.  However, an improved calculation by the same collaboration, employing a larger lattice volume, an extended interpolating basis, and an optimized box-sink construction, found the lowest-lying levels in both the isoscalar $J^P=1^+$ and $0^+$ channels to lie slightly above the corresponding $B^*\olsi D$ and $B\olsi D$ thresholds~\cite{Hudspith:2020tdf}. Similar conclusions were reached in Ref.~\cite{Meinel:2022lzo}, where the extracted energy levels were consistent with the corresponding thresholds, although the statistical uncertainties were too large to rule out the presence of shallow bound states.

More recently, Ref.~\cite{Alexandrou:2023cqg}, combining tetraquark interpolating fields with local and nonlocal meson-meson operators, reported evidence for isoscalar $B\olsi D$ and $B^*\olsi D$ hadronic molecules with $J^P=0^+$ and $1^+$, respectively. This interpretation is supported by studies of the pion-mass dependence of these systems, which suggest that both channels become bound as the pion mass approaches its physical value~\cite{Padmanath:2023rdu,Radhakrishnan:2024ihu}.

Recently, the lattice spectra of the isoscalar $B\olsi D$ and $B^*\olsi D$ systems reported in Ref.~\cite{Alexandrou:2023cqg} were reanalyzed within an effective field theory (EFT) framework~\cite{Abolnikov:2026sik}. The lattice energy levels were described using contact interactions supplemented by one-pion exchange (OPE), with single- and coupled-channel dynamics treated within the Lippmann--Schwinger equation (LSE). Exploiting HQSS, the resulting interaction predicts a family of shallow $B^{(*)}\olsi D{}^{(*)}$ molecular states. An important conclusion of this study is that $S$-wave OPE plays only a minor role in determining the lattice energy levels.

To clarify the role of the OPE interaction, we reanalyze the lattice energy levels using a single-channel treatment of the $B\olsi D$ system and a coupled-channel treatment of the $B^*\olsi D$--$B\olsi D{}^{*}$ system. Following the approach developed in Ref.~\cite{Yu:2025gzg}, we project the effective interactions onto the relevant irreducible representations of the cubic group, namely the $A_1^+$ and $T_1^+$ irreducible representations (irreps) for the $J^P=0^+$ and $1^+$ channels, respectively, thus incorporating the OPE interaction consistently in a finite volume. Using the low-energy constants (LECs) determined from the LQCD spectra of Ref.~\cite{Alexandrou:2023cqg}, we solve the LSE to determine the pole structure of the isoscalar $B^{(*)}\olsi D{}^{(*)}$ systems.
 
This paper is organized as follows. In Sec.~\ref{Sec:Formula}, we present the formalism for the $B^{(*)}\olsi D{}^{(*)}$ scattering amplitudes in both finite and infinite volumes. In Sec.~\ref{Sec:Finite_volume}, we reanalyze the LQCD spectra of the isoscalar $J^P=0^+$ and $1^+$ channels. The resulting predictions for the isoscalar $B^{(*)}\olsi D{}^{(*)}$ hadronic molecules are presented in Sec.~\ref{Sec:Pole}. Finally, we summarize our findings in Sec.~\ref{Sec:Summary}.  
Appendix~\ref{sec:AppendixA} provides details on the partial-wave-projected potentials for the $B^*\olsi D{}^*$ system, while Appendix~\ref{sec:AppendixB} collects the statistical correlation matrices for the $B^*\olsi D$--$B\olsi D{}^*$ system corresponding to the different fits considered in the analysis.

\section{Formalism}\label{Sec:Formula}

In this section, we present the formalism for the scattering of isoscalar $B^{(*)}\olsi D{}^{(*)}$ systems with quantum numbers $J^P=0^+$, $1^+$, and $2^+$ in both infinite and finite volumes. The relevant coupled channels are
\begin{align}
0^+ &: \left\{B\olsi D,\; B^*\olsi D{}^*\right\},\\[3pt]
1^+ &: \left\{B\olsi D{}^*,\; B^*\olsi D,\; B^*\olsi D{}^*\right\},\\[3pt]
2^+ &: \left\{B^*\olsi D{}^*\right\}.
\end{align}
However, since the two-meson $B^*\olsi D{}^*$ scattering operator was not included in the LQCD calculation of Ref.~\cite{Alexandrou:2023cqg}, 
we neglect the coupled-channel effects involving $B^*\olsi D{}^*$, namely the $B\olsi D$--$B^*\olsi D{}^*$ coupling in the $J^P=0^+$ sector and the $(B^*\olsi D,\;B\olsi D{}^*)$--$B^*\olsi D{}^*$ couplings in the $J^P=1^+$ sector.

\subsection{\boldmath $B^{(*)}\olsi D{}^{(*)}$ interactions} \label{subsec:potential}

In the heavy-quark limit, the low-energy interaction between $B^{(*)}\olsi D{}^{(*)}$ mesons is dominated by light degrees of freedom and follows HQSS constraints, which require invariance under separate spin rotations of the $\olsi b$ and $\olsi c$ anti-quarks and allow the corresponding potentials to be constructed systematically for different spin-parity channels~\cite{Nieves:2012tt,Hidalgo-Duque:2012rqv, Bondar:2011ev}. Corrections to HQSS relations are suppressed by powers of $\Lambda_{\rm QCD}/M_Q$ for finite heavy-quark masses, where $M_Q$ denotes the heavy-quark mass. 
For center-of-mass (c.m.) energies close to the relevant $B^{(*)}\olsi D{}^{(*)}$ thresholds, one can construct chiral EFT, similarly to that for the nucleon force~\cite{Epelbaum:2008ga}. At leading order (LO), the interaction consists of constant contact terms and OPE, representing the short- and long-distance dynamics, respectively.

For the isoscalar $B\olsi D\to B\olsi D$ and $B^*\olsi D{}^*\to B^*\olsi D{}^*$ channels with quantum numbers $J^P=0^+$, the LO contact potentials take the form
\begin{align}
V_C^0(B\olsi D)&= C_{0a}, \nonumber\\
V_C^0(B^*\olsi D{}^*)&= C_{0a}-2 C_{0b},
\label{Eq:contact_S0}
\end{align}
where the superscript $J$ in $V_C^J$ denotes the total angular momentum of the system. $C_{0a}$ and $C_{0b}$ are the independent LECs. Similarly, the contact potentials for the isoscalar  $\{B\olsi D{}^*,B^*\olsi D\}$ coupled-channel system and for the $B^*\olsi D{}^{*}$ system are
\begin{align}
 V_C^1(B^*\olsi D,B\olsi D{}^*)&=\begin{pmatrix}
C_{0a} & C_{0b} \\
C_{0b} & C_{0a} \\
\end{pmatrix},\nonumber\\
V_C^1(B^*\olsi D{}^*)&=C_{0a}-C_{0b},\nonumber\\
V_C^2(B^*\olsi D{}^*)&=C_{0a}+C_{0b}.
\label{Eq:contact_S1}
\end{align}

For the long-distance dynamics, the LO Lagrangian describing the $\olsi D{}^{(*)}\olsi D{}^{(*)}\pi$ and $B{}^{(*)} B{}^{(*)}\pi$ couplings in the framework of heavy-meson chiral perturbative theory (ChPT) is~\cite{Wise:1992hn,Yan:1992gz,Burdman:1992gh}
\begin{align}
{\cal L}_{\text{ch}} =& \frac{g}{F_{\pi}}\left(P_a^{\dagger} P_b^{* i} \partial^i \Phi_{b a}+ P_a^{* i \dagger} P_b \partial^i \Phi_{b a} + i \epsilon^{i j k} P_a^{* i \dagger} P_b^{* j} \partial^k \Phi_{b a}\right),
\label{Eq:Lag_chpt_H}
\end{align}
with the pseudoscalar and vector fields $P^{(*)}_a=(\olsi D{}^{(*)0}, D^{(*)-})$ 
and $(B{}^{(*)+}, B{}^{(*)0})$, respectively. The pion fields are collected in
\begin{align}
\Phi=\begin{pmatrix}
\pi^{0}& \sqrt{2}\pi^{+} \\
\sqrt 2\pi^{-} & -\pi^{0}
\end{pmatrix}.
\end{align}
Heavy-quark flavor symmetry (HQFS) implies that the transitions $\olsi D{}^*\to \olsi D{}^{(*)}\pi$ and $B^*\to B^{(*)}\pi$ share the same coupling constant $g$. The measured pionic $D^{*+}$ decay width~\cite{ParticleDataGroup:2026mpi} gives $g=0.57(1)$ in the charm sector, while LQCD with the physical bottom quark yields $g=0.56(3)_{\rm stat}(7)_{\rm sys}$ after chiral extrapolation~\cite{Flynn:2015xna}, consistent within uncertainties.

At the physical pion mass, we use the pion decay constant $F_{\pi}=92.2$ MeV and the effective coupling $g=0.57$. At unphysical pion masses, their dependence on $m_\pi$ is taken into account following Refs.~\cite{Du:2023hlu,Baru:2013rta,Becirevic:2012pf}. For $m_{\pi}=220\,\text{MeV}$, we have $g=0.61$ and $F_{\pi}=99.4\,\text{MeV}$, although the variation on $g^2/F_\pi^2$ with respect to the physical point is less than one percent. 

Defining the isoscalar wave function of $B^{(*)}\olsi D{}^{(*)}$ systems
\begin{align}
\left|I= 0,I_3 = 0\right\rangle = \frac{1}{\sqrt 2}\left(\left|B^{(*)+} D^{(*)-}\right\rangle - \left|B^{(*)0}\olsi D{}^{(*)0}\right\rangle \right),
\end{align}
the OPE potentials for the $B^*\olsi D$--$B\olsi D{}^*$ and $B^*\olsi D{}^*$ systems are,
\begin{align}
V_{\OPE}(B^*\olsi D,B\olsi D{}^*) &=-\frac{3g^2}{4F_{\pi}^2}D_{\pi}(E)
\begin{pmatrix}
& 0 ~~~& [\boldsymbol q\cdot \varepsilon^*(\boldsymbol k_1)] [\boldsymbol q\cdot \varepsilon(\boldsymbol p_2)] \\
& [\boldsymbol q\cdot \varepsilon^*(\boldsymbol k_2)][ \boldsymbol q\cdot \varepsilon(\boldsymbol p_1)]~~~ & 0 ~~~~~ \\
\end{pmatrix},\nonumber\\[3pt]
V_{\OPE}(B^*\olsi D{}^*)&=\frac{3g^2}{4F_{\pi}^2}D_{\pi}(E){\left[\boldsymbol q \cdot\left(\varepsilon^*(\boldsymbol k_1)\times \varepsilon(\boldsymbol p_1)\right)\right] \left[\boldsymbol q \cdot\left(\varepsilon^*(\boldsymbol k_2)\times \varepsilon(\boldsymbol p_2)\right)\right]},
\label{Eq:OPE}
\end{align}
 where $\varepsilon(\boldsymbol p_i)$ denotes the polarization vector of the vector meson, $\boldsymbol p_i$ and $\boldsymbol k_i$ denote the three-momenta of the initial and final states, respectively, with $i=1$ and $2$ for the $B^{(*)}$ and $\olsi D{}^{(*)}$, respectively. The exchanged pion momentum is denoted by $\boldsymbol{q} = \boldsymbol{p}_1-\boldsymbol{k}_1$. The pion propagator $D_{\pi}(E)$ with c.m. energy $E$ can be decomposed into two contributions in terms of the time-ordered perturbation theory (TOPT),
\begin{align}
D_\pi(E)=\frac{1}{2 E_\pi}\left[\frac{1}{E-\omega_2(\boldsymbol p_2)-\omega_1(\boldsymbol k_1)-E_\pi(\boldsymbol q)+i \varepsilon}+\frac{1}{E-\omega_1({\boldsymbol p_1})-\omega_2({\boldsymbol k_2})-E_\pi(\boldsymbol q)+i \varepsilon}\right],
\end{align}
with $\omega_i({\boldsymbol p_i})=\sqrt{m_i^2+ \boldsymbol p_i^2}$ and $E_{\pi}(\boldsymbol q)=\sqrt{m_{\pi}^2+ \boldsymbol q^2}$.

The OPE potentials are decomposed into the different partial-wave components using the partial-wave projection method~\cite{Gulmez:2016scm,Shi:2024squ}. In the c.m. frame, the three momenta of the initial and final states are chosen as
\begin{align}
 \boldsymbol p_1=-\boldsymbol p_2=(0,0,p),\quad
 \boldsymbol k_1=-\boldsymbol k_2=(k\,\text{sin}\,\theta \,\text{cos}\,\phi,k\,\text{sin}\,\theta \,\text{sin}\,\phi,k\,\text{cos}\,\theta).
 \end{align}
Then, the OPE potential in the isoscalar $B^*\olsi D({\cal S})$--$B^*\olsi D({\cal D})$--$B\olsi D{}^*({\cal S})$--$B\olsi D{}^*({\cal D})$ coupled-channel system is given in the partial-wave basis by
\begin{align}
V^1_{\OPE}(\boldsymbol{q}, B^*\olsi D,B\olsi D{}^*)=-\frac{1}{8}\frac{g^2}{F_{\pi}^2}
\begin{pmatrix}
& 0 & 0 & V_{\cal SS}(p,k) & V_{\cal SD}(p,k)  \\
& 0 & 0 & V_{\cal DS}(p,k) & V_{\cal DD}(p,k)  \\
& V_{\cal SS}(p,k) & V_{\cal SD}(p,k) & 0  & 0 \\
& V_{\cal DS}(p,k) & V_{\cal DD}(p,k) & 0  & 0 \\
\end{pmatrix},
\label{Eq:pot_partial_wave}
\end{align}
where $\cal S$ and $\cal D$ denote $S$- and $D$-wave components, respectively. The partial-wave elements are defined as
\begin{align}
V_{\cal SS}(p,k)&\equiv\int_{-1}^1 \text{d}z\, D_{\pi}(E)\left({p^2+k^2-2pkz}\right),\nonumber\\
V_{\cal SD}(p,k)&\equiv\int_{-1}^1 \text{d}z\, D_{\pi}(E)\left[{(p^2-2k^2)+4pkz-3p^2z^2}\right]/\sqrt{2},\nonumber\\
V_{\cal DS}(p,k)&\equiv\int_{-1}^1 \text{d}z\, D_{\pi}(E)\left[{(k^2-2p^2)+4pkz-3k^2z^2}\right]/\sqrt{2},\nonumber\\
V_{\cal DD}(p,k)&\equiv\int_{-1}^1 \text{d}z\, D_{\pi}(E)\left[{(k^2+p^2)(3z^2-1)+pkz(1-9z^2)/2}\right].
\label{Eq:partial_wave_component}
\end{align}
Here, $z=(\boldsymbol p_1 \cdot \boldsymbol k_1)/pk$ denotes the cosine of the angle between the initial and final relative momenta.
The corresponding partial-wave OPE potentials for the $B^*\olsi D{}^*$ system are detailed in Appendix~\ref{sec:partial_wave}. 

\subsection{\boldmath $T$-matrix in the infinite and finite volumes}

In the infinite volume, the scattering of the coupled-channel system can be described by the $T$-matrix
\begin{align}
T_{\eta\delta}(E,\boldsymbol p,\boldsymbol k) = \sum_{\alpha\beta}F^{\Lambda}_{\eta\alpha}(\boldsymbol p) T^{\prime}_{\alpha\beta}(E,\boldsymbol p,\boldsymbol k) F^{\Lambda}_{\beta\delta}(\boldsymbol k),
\label{Eq:T-matrix-coupled}
\end{align} 
with
\begin{align}
T^{\prime}_{\alpha\beta}(E,\boldsymbol p,\boldsymbol k)=V_{\alpha\beta}(E,\boldsymbol p,\boldsymbol k) +\sum_{\gamma}\int\frac{d\boldsymbol q^3}{(2\pi)^3}V_{\alpha\gamma}(E,\boldsymbol p,\boldsymbol{q})I_{\gamma\gamma}^{\Lambda}(E,\boldsymbol{q}) T^{\prime}_{\gamma\beta}(E,\boldsymbol{q},\boldsymbol k).\label{Eq:T-matrix-no-factor}
\end{align}
Here, the Greek subscripts label the channels involved in the system, and $\boldsymbol p$ and $\boldsymbol k$ denote the three-momenta of initial and final states, respectively. A Gaussian regulator is employed to render the Green's function ultraviolet  (UV) finite,
\begin{align}
I_{\alpha\beta}^{\Lambda}(E,\bq)=\frac{\delta_{\alpha\beta}\, e^{-2\boldsymbol \bq^2/\Lambda^2}}{E-(\omega_1(\bq)+\omega_2(\bq))+i\varepsilon},
\end{align}
and correspondingly a form factor $F^{\Lambda}_{\eta\alpha}(\boldsymbol p)=\delta_{\eta\alpha}\, e^{-\boldsymbol p^2/\Lambda^2}$ is introduced to ensure the coupled-channel unitarity under this regularization~\cite{Cincioglu:2016fkm,Shi:2024llv}.
In the $B^*\olsi D$--$B\olsi D{}^*$ coupled-channel system, the relative momentum of the $B^*\olsi D$ pair is approximately $0.5$~GeV when the c.m.\,energy is at the $B\olsi D{}^*$ threshold. Accordingly, the cutoff value needs to be larger, and we adopt  $\Lambda=0.8$ and $1.2$ GeV to check the dependence of the results on the UV regulator. 
The total effective potentials comprise both contact and OPE contributions,
\begin{align}
V(E,\boldsymbol p,\boldsymbol k)=V_C + V_{\OPE}(\boldsymbol p,\boldsymbol k),
\end{align}
where the $V_C$ and $V_{\OPE}$ for different angular momentum $J$ correspond to the formulas given in Eqs.~\eqref{Eq:contact_S0}, \eqref{Eq:contact_S1}, and \eqref{Eq:OPE}.

Under analytic continuation, the complex energy plane is divided into different Riemann sheets (RSs). Neglecting the decay widths of the heavy mesons, the two-body branch cut lies along the real axis in the complex $E$-plane, allowing the different RSs to be characterized by a straightforward relation
\begin{align}
\left[ T^{(\sigma_1\sigma_2)}(E)^{-1} \right]_{\alpha\beta} =
\left[ T^{(++)}(E)^{-1} \right]_{\alpha\beta} + 2i\,\frac{\mu_{\alpha}\,p_\alpha(E)}{2\pi}\,\delta_{\alpha\beta}\,\theta_{\sigma_\alpha}\,,
\end{align}
where $T_{\alpha\beta}(E)\equiv T_{\alpha\beta}\big(E,p_\alpha(E),p_\beta(E)\big)$ is the on-shell $T$-matrix, $p_\alpha(E)$ is the c.m.\ on-shell momentum of channel $\alpha$, analytically continued from the physical region (${\rm Re}\,p_\alpha>0$, ${\rm Im}\,p_\alpha\le 0$ for ${\rm Im}\,E\le 0$), and $\mu_\alpha$ is the reduced mass of channel $\alpha$. The RSs are labeled by $(\sigma_1\sigma_2)$, with $\sigma_\alpha=+$ $(-)$ when channel $\alpha$ is on its first (second) sheet, and $\theta_{\sigma_\alpha}=1$ if $\sigma_\alpha=-$ and $0$ otherwise. Partial waves of the same two-meson channel share $p_\alpha$ and $\sigma_\alpha$. The pole positions of $B^{(*)}\olsi D{}^{(*)}$ molecules, $E_R$, can be searched for on each RS. In addition, the effective coupling of a molecular state to the $\alpha$-channel $g_\alpha$ is extracted from the residue of the $T$-matrix evaluated on the relevant RS 
\begin{align}
g_{\alpha}g_{\beta}=\lim_{E\to E_R}(E-E_R)T_{\alpha\beta}(E).
\end{align}

In finite volume, the standard L\"uscher formula can be used to extract scattering information from lattice data as long as the left-hand cut (LHC) lies below the energy region of interest. However, when the energy levels lie close to the left-hand cut, the extraction of scattering information becomes more challenging~\cite{Du:2023hlu}. An effective way to overcome this difficulty is to explicitly incorporate the long-range interaction responsible for the left-hand cut in the finite-volume formalism. To analyze the LQCD energy levels, the long-range interaction must then be projected onto the relevant irreducible representations (irreps) of the cubic group. In this work, we follow the approach of Ref.~\cite{Yu:2025gzg} to reduce the effective potentials into a specific irrep. For a vector--pseudoscalar meson system, the finite-volume $T^{\prime}(E,\boldsymbol p,\boldsymbol k)$ in Eq.~\eqref{Eq:T-matrix-no-factor} is rewritten as
\begin{align}
\tilde T^{\prime\,\Gamma}_{\alpha\lambda_1 r_1,\beta\lambda_2 r_2}&(E,\boldsymbol p,\boldsymbol k) = \tilde V^{\Gamma}_{\alpha\lambda_1 r_1,\beta\lambda_2 r_2}(E,\boldsymbol p,\boldsymbol k) \nonumber\\
&+\sum_{\gamma,\lambda_3,r_3,\bqn} \tilde V^{\Gamma}_{\alpha\lambda_1r_1,\gamma\lambda_3r_3}(E,\boldsymbol p,\bqn)\tilde I^{\Lambda}_{\gamma\lambda_3r_3,\gamma\lambda_3r_3}(E,\bqn)\tilde T^{\prime\,\Gamma}_{\gamma\lambda_3r_3,\beta\lambda_2r_2}(E,\bqn,\boldsymbol k),
\label{Eq:T_matrix_FV}
\end{align}
where $\lambda_i$ and $r_i$ with $i=1,\,2$ and $3$ denote the helicity of the vector meson and the row of irrep $\Gamma$, respectively.\footnote{In a system with two pseudoscalar mesons, we neglect the helicity indices in the $T$-matrix of Eq.~\eqref{Eq:T_matrix_FV}, while adding other relevant helicity indices for a system with two vector mesons.} Given a size of finite volume $L$, the finite-volume Green's function is
\begin{align}
\tilde I^{\Lambda}_{\alpha\lambda_1r_1,\beta\lambda_2r_2}(E,\bqn)=\frac{1}{L^3}\frac{\delta_{\alpha\beta}\delta_{\lambda_1\lambda_2}\delta_{r_1r_2}e^{-2\bqn^2/\Lambda^2}}{E-\tilde\omega_{B^{(*)}}(\bqn)-\tilde\omega_{\olsi D{}^{(*)}}(\bqn)},
\end{align}
where $\tilde\omega_{B^{(*)}}(\bqn)$ and $\tilde\omega_{\olsi D{}^{(*)}}(\bqn)$ satisfy the dispersion relations of $B^{(*)}$ and $\olsi D{}^{(*)}$ mesons in Ref.~\cite{Alexandrou:2023cqg}. 

The potential $\tilde V^{\Gamma}_{\alpha\lambda_1 r_1,\beta\lambda_2 r_2}(E,\boldsymbol p,\boldsymbol k)$ is calculated using the projection operator 
\begin{align}
\tilde V^{\Gamma}_{\alpha\lambda_1 r_1,\beta\lambda_2 r_2}(E,\boldsymbol p,\boldsymbol k)=\langle \boldsymbol p,\Gamma,\lambda_1,r_1,\mu|\hat V_{\alpha\beta}(E,\boldsymbol p,\boldsymbol k)|\boldsymbol k,\Gamma,\lambda_2,r_2,\mu \rangle\,,
\label{Eq:Potential_FV}
\end{align}
and we note that the matrix element does not depend on $\mu$. Here the orthonormal basis of the $\Gamma$ irrep is 
\begin{align}
|\boldsymbol p,\Gamma,\lambda,r, \mu \rangle = N^{\Gamma}_{\boldsymbol p\lambda}\sum_{\nu} \left[c_{\boldsymbol p,\lambda}^{\Gamma,r}\right]_\nu P_{\mu\nu}^{\Gamma} |\boldsymbol p,\lambda \rangle,
\end{align}
where $N^{\Gamma}_{\boldsymbol p\lambda}$ and $c_{\boldsymbol p,\lambda}^{\Gamma,r}$ denote the normalization factor and the $r$-th orthonormal eigenvector of the $\Gamma$ irrep, respectively; see details in Ref.~\cite{Yu:2025gzg}. The projection operator is defined as
\begin{align}
P_{\mu \nu}^{\Gamma}=\frac{\text{dim}(\Gamma)}{|G|} \sum_{g \in G} \olsi{D}{}_{\mu \nu}^{\Gamma}(g) g,
\end{align}
with the cubic group $G$ and the matrix of the $\Gamma$ irrep $\olsi{D}{}^{\Gamma}(g)$. 

The LECs $C_{0a}$ and $C_{0b}$ are determined from fitting the lattice data for the $\bar b \bar c ud$ system~\cite{Alexandrou:2023cqg}. For the $B\olsi D$ system in the $A_1^+$ irrep, the finite-volume potential reduces to the contact potential of Eq.~\eqref{Eq:contact_S0}, since long-range interactions do not contribute in the single-channel formalism. In contrast, for the coupled-channel $B^*\olsi D$--$B\olsi D{}^*$ system, the full potential, $V=V_C^1(B^*\olsi D,B\olsi D{}^*)+V_{\OPE}(B^*\olsi D,B\olsi D{}^*)$, must be projected onto the $T_1^+$ irrep.

\section{\boldmath Reanalysis of the energy levels for $\bar b \bar c ud$ systems}\label{Sec:Finite_volume}

As mentioned above, the LECs entering the effective potentials are determined by reanalyzing the $B \olsi D$ and $B^* \olsi D$ energy levels reported in Ref.~\cite{Alexandrou:2023cqg}. The LQCD calculations were performed on two lattice volumes, $L=24a$ and $32a$, with a weighted-average lattice spacing of $a\simeq 0.11887(80)$ fm and a pion mass of $m_{\pi}\simeq 220$ MeV. For the isoscalar $J^P=0^+$ $B\olsi D$ system, four and five energy levels are reported below the $B^* \olsi D{}^*$ threshold for the two lattice volumes. These levels are extracted using local and nonlocal $B\olsi D$ interpolating operators, together with local $B^*\olsi D{}^*$-type and tetraquark operators. For the $J^P=1^+$ system, five energy levels are obtained for each lattice volume using local operators of the $B^*\olsi D$, $B\olsi D{}^*$, and $B^*\olsi D{}^*$ types, as well as tetraquark operators, supplemented by nonlocal $B^*\olsi D$ scattering operators. 
Among these energy levels, the third one is dominated by the $D$-wave component, as it couples strongly only to operators subduced from a $D$-wave structure~\cite{Alexandrou:2023cqg}. Here, ``local'' refers to operators in which the quark fields are combined at the same spacetime point and projected onto zero momentum, whereas ``nonlocal'' scattering operators correspond to back-to-back momentum configurations in the rest frame. 
The local $B\olsi D^{*}$ and $B^*\olsi D$ operators are both included in the simulation of the $B^*\olsi D$ system, as well as local $\bar{b} \bar{c} u d$ operators that can be rearranged into combinations of meson-meson operators. 
We reanalyze the energy levels in the $T_1^+$ irrep using the $B^*\olsi D$--$B\olsi D{}^{*}$ coupled-channel formalism. However, we exclude the energy levels close to the $B\olsi D{}^*$ threshold in our fit due to the lack of nonlocal $B\olsi D{}^*$ operators, which makes a direct comparison with our coupled-channel description less clear in this energy region. 
For the $A_1^+$ irrep, the $\chi^2$ dramatically increases when the third energy level is included in the fit. This means that the LO contact potentials fail to describe the higher energy levels. Consequently, we determine the LECs by performing a simultaneous fit to the first two energy levels in the $A_1^+$ and $T_1^+$ irreps. It is worth noting that the omission of the higher energy levels has only a negligible effect on the binding energies of the $B\olsi D$ and $B^*\olsi D$ molecules.\footnote{In Ref.~\cite{Alexandrou:2023cqg}, the binding energies extracted from the full spectrum were compared with those obtained using only the lowest three energy levels combined across both lattice volumes. For both the $A_1^+$ and $T_1^+$ irreps, the differences remain within $1\sigma$.}

To investigate the impact of the OPE potential, we reanalyze the complete set of eight energy levels using different models and cutoff values:
\begin{itemize}
\item Model IA (IB): Only the contact potentials in Eqs.~\eqref{Eq:contact_S0} and \eqref{Eq:contact_S1} are included for the $B\olsi D$ and $B^*\olsi D$--$B\olsi D{}^*$ systems, with a cutoff of $\Lambda=0.8$ (1.2) GeV.

\item Model IIA (IIB): The contact potential in Eq.~\eqref{Eq:contact_S0} is included for the $B\olsi D$ system (notice that a single pion cannot be exchanged between $B\olsi D$ without coupled channels of higher thresholds), whereas both the contact and OPE potentials in Eqs.~\eqref{Eq:contact_S1} and \eqref{Eq:OPE} are included for the coupled-channel $B^*\olsi D$--$B\olsi D{}^*$ system, with a cutoff of $\Lambda=0.8$ ($1.2$) GeV.
\end{itemize}
The finite-volume energy levels are determined from the poles of the $T$-matrix in Eq.~\eqref{Eq:T_matrix_FV}. The lattice energy levels from Ref.~\cite{Alexandrou:2023cqg} are then fitted by minimizing the corresponding $\chi^2$ function using the MINUIT minimization package~\cite{James:1975dr,iminuit,iminuit.jl}.

The fitted LECs for the different models are summarized in Table~\ref{Tab:LECs}. For both cutoff values, $\Lambda=0.8$ and $1.2$~GeV, the values of $C_{0a}$ obtained in Models I and II differ by less than $0.5\sigma$, whereas the corresponding values of $C_{0b}$ differ by about $2\sigma$. This difference is expected because the contact terms also act as counterterms that absorb short-distance UV divergences and are therefore cutoff dependent. In particular, Eqs.~\eqref{Eq:contact_S0} and \eqref{Eq:contact_S1} show that the $C_{0a}$-only matrix elements correspond to processes for which OPE is absent, whereas $C_{0b}$ also renormalizes the OPE interaction at LO.

\begin{table}[t]
\caption{\label{Tab:LECs} LECs determined using various models (see main text) by fitting the lattice energy levels reported in Ref.~\cite{Alexandrou:2023cqg}. }
\renewcommand{\arraystretch}{1.2}
\begin{tabular*}{\columnwidth}{@{\extracolsep\fill}lcccc}
\hline\hline 
 Model & Model IA & Model IB & Model IIA & Model IIB \\[3pt]     
\hline
     $C_{0a}$ [fm$^2$] & $-0.71_{-0.04}^{+0.04}$  & $-0.40_{-0.02}^{+0.02}$ & $-0.71_{-0.04}^{+0.04}$
     & $-0.39_{-0.01}^{+0.01}$ \\[3pt]
     $C_{0b}$ [fm$^2$]  & $0.31_{-0.11}^{+0.09}$  & $0.12_{-0.05}^{+0.04}$  & $0.00_{-0.12}^{+0.09}$ & $-0.27^{+0.07}_{-0.09}$   \\[3pt]
     $\chi^2/\text{d.o.f.}$ & $7.66/6$  & $1.08/6$  & $7.77/6$ & $2.09/6$  \\[3pt]
\hline\hline
\end{tabular*}
\end{table}

To further investigate the cutoff dependence of the LECs, we vary $\Lambda$ from $0.6$ to $1.2\,\text{GeV}$ and examine the resulting evolution of the fitted LECs, as shown in Fig.~\ref{Fig:LEC_cutof}. For $\Lambda=0.6\,\text{GeV}$, the resulting $\chi^2$ becomes unacceptably large. This is not surprising, since such a small cutoff is close to the $B^*\olsi D$ momentum ($\sim0.5\,\text{GeV}$) at the $B\olsi D{}^*$ threshold and is therefore insufficient to properly describe the wave function of the $B^{(*)}\olsi D{}^{(*)}$ system~\cite{Albaladejo:2017blx}. Furthermore, as shown in Fig.~\ref{Fig:LEC_cutof}, once the OPE potential is included, the value of $C_{0b}$ is smaller than that obtained with the contact potential alone, which reflects the consistent absorption of the short-distance part of the OPE interaction into the corresponding counterterms.

\begin{figure*}[t]
     \centering	
     \includegraphics[width=0.46\textwidth]{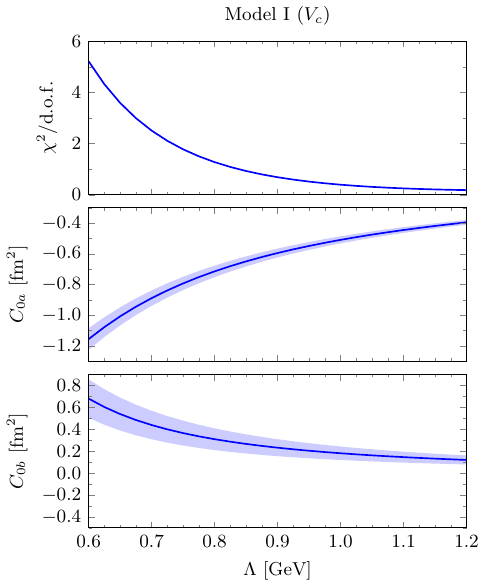}
     \includegraphics[width=0.46\textwidth]{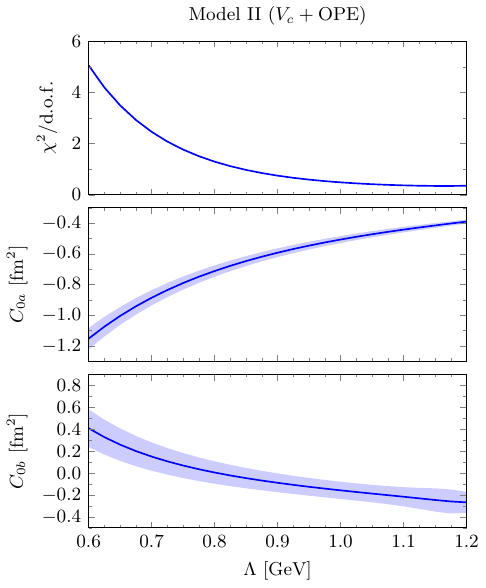}
     \caption{Cutoff dependence of $\chi^2/\text{d.o.f.}$ (top panels) and of the LECs $C_{0a}$ (middle panels) and $C_{0b}$ (bottom panels), extracted from fits to the LQCD energy levels reported in Ref.~\cite{Alexandrou:2023cqg}. The left and right columns show the results for Model I, which includes only the contact potential, and Model II, which includes both the contact and OPE potentials, respectively. The bands indicate the $1\sigma$ uncertainties estimated from the Hessian of the $\chi^2$ function.\label{Fig:LEC_cutof}}
\end{figure*}

In a finite volume, the breaking of Lorentz symmetry induces mixing among different partial waves. As the cutoff $\Lambda$ increases, higher-partial-wave contributions generated by the long-range interaction become increasingly important, leading to a more pronounced modification of the fitted value of $C_{0b}$. A detailed discussion of the impact of higher partial waves on the phase shifts induced by the long-range interaction can be found in Ref.~\cite{Yu:2025gzg}.

\begin{figure*}
     \centering	
     \includegraphics[width=0.45\textwidth]{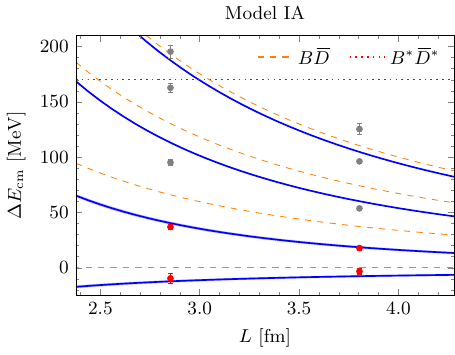}
     \includegraphics[width=0.45\textwidth]{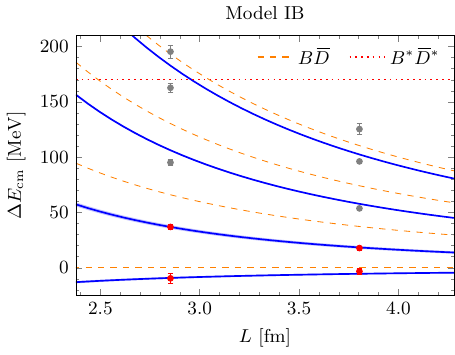}
     \includegraphics[width=0.45\textwidth]{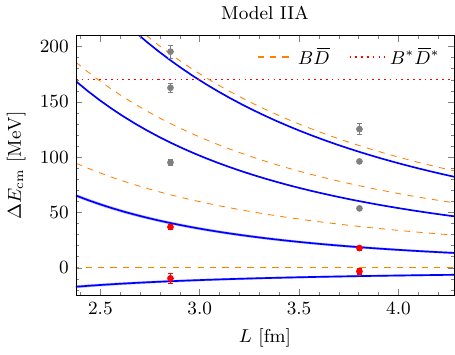}
     \includegraphics[width=0.45\textwidth]{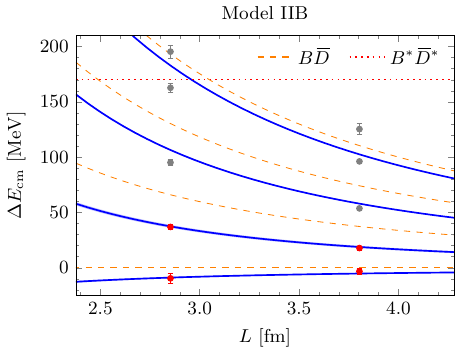}
     \caption{Finite-volume energy levels of the $B\olsi D$ system computed using the LECs listed in Table~\ref{Tab:LECs}. The points denote the lattice energy levels reported in Ref.~\cite{Alexandrou:2023cqg} in the $A_1^+$ irrep, with the red (gray) points indicating the levels included (excluded) in the fits. The dashed lines represent the non-interacting energy levels of the $B\olsi D$ pair, while the dotted line denotes the $B^*\olsi D^*$ threshold. The blue bands correspond to the $68\%$ confidence intervals propagated from the uncertainties of the fitted LECs. 
     \label{Fig:EL_0}}
\end{figure*}
     
\begin{figure*}
     \centering	
     \includegraphics[width=0.45\textwidth]{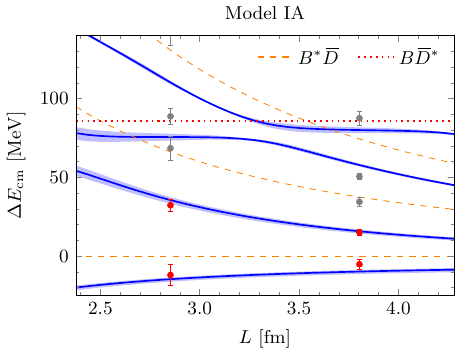}
     \includegraphics[width=0.45\textwidth]{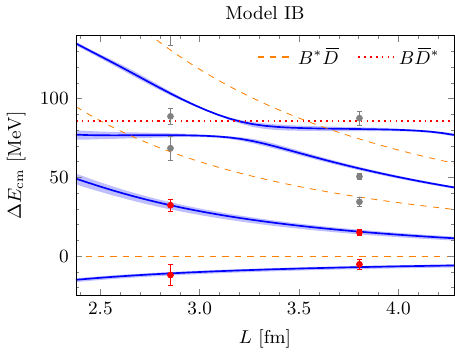}\\
     \includegraphics[width=0.45\textwidth]{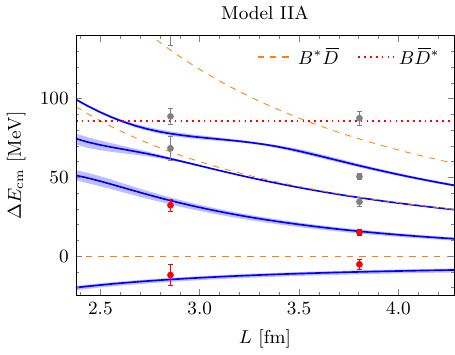}
     \includegraphics[width=0.45\textwidth]{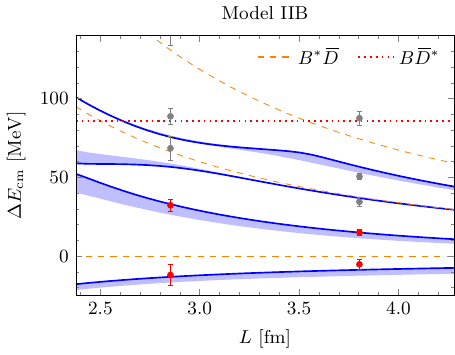}
     \caption{Same as Fig.~\ref{Fig:EL_0}, but for the $B^*\olsi D$--$B\olsi D{}^*$ coupled-channel system and the levels in the $T_1^+$ irrep. Here the dashed lines represent the non-interacting energy levels of the $B^*\olsi D$ pair, while the dotted line denotes the $B\olsi D^*$ threshold. 
     \label{Fig:EL_1}}
\end{figure*}

Using the LECs listed in Table~\ref{Tab:LECs}, we compute the volume dependence of the energy levels, as shown in Figs.~\ref{Fig:EL_0} and~\ref{Fig:EL_1}. In the $A_1^+$ irrep, the lowest energy level remains below the $B\olsi D$ threshold even at the largest volume considered, providing clear evidence for a $B\olsi D$ bound state. However, our analysis fails to reproduce the third and higher energy levels, 
As discussed in Ref.~\cite{Alexandrou:2023cqg}, a more accurate description of these higher-lying levels is expected to require the inclusion of higher-order terms in the effective-range expansion, as well as the contribution from the $ B^*\olsi D{}^*$ channel in the L\"uscher analysis~\cite{Abolnikov:2026sik}. More generally, analyses of lattice spectra for other heavy--light systems have shown that going beyond the LO contact interaction, either by including next-to-leading-order terms in the potential or by explicitly incorporating $s$-channel states, can be useful~\cite{Albaladejo:2016lbb,Albaladejo:2016ztm,Albaladejo:2018mhb,Albaladejo:2026yfz}.

In the $T_1^+$ irrep, all four models predict a plateau near the $B\olsi D{}^*$ threshold, signaling a narrow resonance in this energy region~\cite{Wiese:1988qy}. Moreover, once the OPE potential is included, Models IIA and IIB reproduce the third energy level observed in the lattice simulation at both volumes, $L=24a$ and $32a$, whereas Models IA and IB do not. In particular, for $L=32a$, the predicted third energy level coincides with the lattice result and lies below the corresponding non-interacting $B^*\olsi D$ energy level with $n=1$.

By contrast, when the $p^2$ contact interaction, with $p$ the c.m. momentum magnitude, is included instead~\cite{Abolnikov:2026sik}, the third energy level also emerges but remains above the corresponding non-interacting level, even when the $S$-wave OPE potential is included. This qualitative difference highlights the importance of the $D$-wave contribution induced by the OPE potential, consistent with the discussion in Ref.~\cite{Alexandrou:2023cqg}. Furthermore, the absence of nonlocal $B\olsi D{}^*$ scattering operators in the lattice calculation makes it difficult to assess the predicted energy levels near the $B\olsi D{}^*$ threshold, as the corresponding lattice spectrum could be significantly modified by their inclusion.

\section{\boldmath Molecular states for the $\bar b \bar c ud$ systems}\label{Sec:Pole}

To predict the properties of the physical hadronic molecules from the LECs extracted from the lattice data, the pion-mass dependence of the effective potentials must be incorporated. In particular, the pion-mass dependence of the LECs is parametrized as~\cite{Baru:2013rta}
\begin{align}
C^{T}_{0a/0b}=C_{0a/0b}+\delta C_{0a/0b},
\end{align}
with
\begin{align}
\frac{\delta C_{0a/0b}}{C_{0a/0b}}\sim (\xi^2-1)\frac{(m_{\pi}^{\text{phys}})^2}{M^2},
\end{align}
where $\xi=m_{\pi}/m_{\pi}^{\text{phys}}$ is the ratio between the unphysical and physical pion masses, and the hard scale is estimated as $M\sim m_{\sigma}\sim 500$~MeV. For $m_{\pi} \simeq 220\,\text{MeV}$, we have $\delta C_{0a/0b}/C_{0a/0b} \simeq 0.12$.
The term $\delta C_{0a/0b}$ is numerically suppressed, so we neglect the pion-mass dependence of the contact potentials. The pion-mass dependence of the OPE potential, however, is already accounted for in our reanalysis of the lattice data. We therefore use the values of $C_{0a}$ and $C_{0b}$ listed in Table~\ref{Tab:LECs}, together with the physical heavy-meson and pion masses, to predict the molecular states in the $\bar b \bar c ud$ system.

In infinite volume, the $S$- and $D$-wave OPE interactions, together with the contact interactions, are used to investigate the possible hadronic molecular states by solving the coupled-channel LSE of Eq.~\eqref{Eq:T-matrix-coupled}. The resulting pole positions of the isoscalar $\bar b \bar c ud$ systems for the different sets of LECs listed in Table~\ref{Tab:LECs} are summarized in Tables~\ref{Tab:pole_IA}--\ref{Tab:pole_IIB}. Note that, on the first RS, the left-hand cut of the scattering amplitude originates solely from the tree-level potential, i.e., the first term in Eq.~\eqref{Eq:T-matrix-no-factor}. The integral term, i.e., the second term, generates the left-hand cut only on RSs other than the first~\cite{Du:2024gzw}. Consequently, bound-state poles may occur below the branch point of the left-hand cut, whereas virtual-state poles on the real axis of an unphysical RS below this branch point are generally forbidden.

\begin{table}[tb]
\caption{\label{Tab:pole_IA} Predicted pole positions of the $\bar b \bar c ud$ systems in Model IA. The effective couplings $g_{\mathrm{low}}$ and $g_{\mathrm{high}}$ denote the couplings of each pole to the lower and higher $S$-wave channels, respectively. For single-channel systems, only $g_{\mathrm{low}}$ is relevant. The $D$-wave couplings are omitted for both single- and coupled-channel systems. The ``Thr.'' column lists the corresponding $H\olsi H$ thresholds calculated using the physical isospin-averaged meson masses. The quoted uncertainties in the pole positions and effective couplings correspond to the $68\%$ confidence intervals obtained by propagating the uncertainties of the fitted LECs using the bootstrap method. The Riemann sheet (RS) on which each pole is located is also indicated.
}
\renewcommand{\arraystretch}{1.2}
\begin{tabular*}{\columnwidth}{@{\extracolsep\fill}lcccccc}
\hline\hline 
 $J^{PC}$ &  $H\olsi H$ & $M$ [MeV] & RS  & $g_{\text{low}}$ [GeV$^{-1/2}$] & $g_{\text{high}}$ [GeV$^{-1/2}$] & Thr. [MeV] \\[3pt]     
\hline
$0^{+}$  & $B\olsi D$ & $7144^{+1}_{-1}$
     & $(+)$ & $0.63^{+0.08}_{-0.08}$ & - & 7147 \\[3pt]
\hline
\multirow{2}{*}{$1^{+}$}  & \multirow{2}{*}{$B^*\olsi D$--$B\olsi D{}^*$} & {$7187^{+1}_{-2}$}
     & $(+,+)$ & {$0.80^{+0.09}_{-0.08}$}  & $ -0.74_{-0.29}^{+0.2
     7}$  & 7192 \\[3pt]
  & & {$7287^{+3}_{-2} -2^{+1}_{-1}i$}
     & $(-,+)$ & {$0.08^{+0.01}_{-0.02}+0.11^{+0.06}_{-0.04}i$} & {$-0.55_{-0.15}^{+0.15}{-}0.17_{-0.16}^{+0.10}i$} & 7288  \\[3pt]
\hline
$0^{+}$  & $B^*\olsi D{}^*$ & {$7293^{+14}_{-16}$}
     & $(+)$ & {$2.23^{+0.62}_{-0.54}$} & - & 7333 \\[3pt]
\hline
$1^{+}$  & $B^*\olsi D{}^*$ & {$7314^{+5}_{-6}$}
     & $(+)$ & {$1.41^{+0.22}_{-0.21}$} & - & 7333 \\[3pt]
\hline
$2^{+}$  & $B^*\olsi D{}^*$ & {$7331^{+~2}_{-11}$} & $(-)$ & $0.41^{+0.12}_{-0.94} i$ & - & 7333 \\[3pt]
\hline\hline
\end{tabular*}
\end{table}

\begin{table}[tb]
\caption{\label{Tab:pole_IB} Same as Table~\ref{Tab:pole_IA}, but for Model IB. 
}
\renewcommand{\arraystretch}{1.2}
\begin{tabular*}{\columnwidth}{@{\extracolsep\fill}lcccccc}
\hline\hline 
 $J^{PC}$ &  $H\olsi H$ & $M$ [MeV] & RS  & $g_{\text{low}}$ [GeV$^{-1/2}$] & $g_{\text{high}}$ [GeV$^{-1/2}$] & Thr. [MeV] \\[3pt]     
\hline
$0^{+}$  & $B\olsi D$ & $7146^{+0}_{-1}$
     & $(+)$ & $0.39^{+0.09}_{-0.12}$ & - & 7147 \\[3pt]
\hline
\multirow{2}{*}{$1^{+}$}  & \multirow{2}{*}{$B^*\olsi D$--$B\olsi D{}^*$} & {$7190^{+1}_{-1}$}
     & $(+,+)$ & $0.55^{+0.09}_{-0.08}$  & ${-}0.34_{-0.15}^{+0.14}$  & 7192 \\[3pt]
  & & {$7288^{+2}_{-2} -1^{+1}_{-1}i$}
     & $(-,+)$ & {$0.03_{-0.02}^{+0.01}+0.13_{-0.05}^{+0.07}i$} & {${-}0.42_{-0.10}^{+0.07}{-}0.19_{-0.15}^{+0.12
     }i$} & 7288  \\[3pt]
\hline
$0^{+}$  & $B^*\olsi D{}^*$ & {$7297^{+16}_{-16}$}
     & $(+)$ & {$1.56^{+0.34}_{-0.36}$} & - & 7333 \\[3pt]
\hline
$1^{+}$  & $B^*\olsi D{}^*$ & {$7319^{+5}_{-5}$}
     & $(+)$ & {$1.04^{+0.15}_{-0.16}$} & - & 7333 \\[3pt]
\hline
$2^{+}$  & $B^*\olsi D{}^*$ & {$7329^{+4}_{-9}$}
     & $(-)$ & {$0.49^{+0.10}_{-1.09}i$} & - & 7333 \\[3pt]
\hline\hline
\end{tabular*}
\end{table}

In the $I(J^P)=0(0^+)$ sector, two molecular states are predicted. The first one lies close to the $B\olsi D$ threshold and is identified as a $B\olsi D$ bound state. Its binding energy is about 3~MeV in Models IA and IIA, and about 1~MeV in Models IB and IIB. Within uncertainties, these values are consistent with the lattice determination in Ref.~\cite{Alexandrou:2023cqg}, as well as with predictions based on one-boson-exchange potentials and HQSS~\cite{Li:2012ss,Sakai:2017avl,Abreu:2022sra}.

The second pole is located below the $B^*\olsi D{}^*$ threshold by approximately 41, 36, 40, and 15~MeV in Models IA, IB, IIA, and IIB, respectively. For these four models, this pole lies about $146$--$172$~MeV above the $B\olsi D$ threshold. Taking into account the uncertainty, its mass is in agreement with the resonance reported in the $J^P=0^+$ $\bar b \bar c ud$ system in Ref.~\cite{Alexandrou:2023cqg}. Although its decay width cannot be determined within the present single-channel analysis, its mass relative to the $B\olsi D$ threshold suggests that the $J^P=0^+$ state observed in Ref.~\cite{Alexandrou:2023cqg} is predominantly a $B^*\olsi D{}^*$ molecular state.
However, its binding energy predicted by Models IA, IB, and IIB is significantly larger than those predicted in Refs.~\cite{Sakai:2017avl,Abreu:2022sra,Abolnikov:2026sik}.

\begin{table}[tb]
\caption{\label{Tab:pole_IIA}  Same as Table~\ref{Tab:pole_IA}, but for Model IIA.
}
\renewcommand{\arraystretch}{1.2}
\begin{tabular*}{\columnwidth}{@{\extracolsep\fill}lcccccc}
\hline\hline 
 $J^{PC}$ &  $H\olsi H$ & $M$ [MeV] & RS  & $g_{\text{low}}$ [GeV$^{-1/2}$] & $g_{\text{high}}$ [GeV$^{-1/2}$] & Thr. [MeV] \\[3pt]     
\hline
$0^{+}$  & $B\olsi D$ & $7144^{+1}_{-1}$
     & $(+)$ & $0.62^{+0.08}_{-0.08}$ & - & 7147 \\[3pt]
\hline
\multirow{2}{*}{$1^{+}$}  & \multirow{2}{*}{$B^*\olsi D$--$B\olsi D{}^*$} & $7186^{+1}_{-2}$
     & $(+,+)$ & $0.84^{+0.09}_{-0.08}$ 
     & $-0.85_{-0.31}^{+0.28}$ 
     & 7192 \\[3pt]
  & & $7286^{+5}_{-3}-6^{+1}_{-1}i$ 
     & $(-,+)$ & $0.10^{+0.02}_{-0.02}+0.14^{+0.06}_{-0.05}i$ 
     & $-0.66_{-0.14}^{+0.16}-0.42_{-0.13}^{+0.10}i$
     & 7288  \\[3pt]
\hline
$0^{+}$  & $B^*\olsi D{}^*$ & $7293^{+14}_{-15}$
     & $(+)$ & $2.34^{+0.61}_{-0.54}$
     & - & 7333 \\[3pt]
\hline
$1^{+}$  & $B^*\olsi D{}^*$ & $7313^{+6}_{-6}$
     & $(+)$ & $1.53^{+0.24}_{-0.24}$ 
     & - & 7333 \\[3pt]
\hline
$2^{+}$  & $B^*\olsi D{}^*$ & $7333^{+0}_{-1}$
     & $(+)$ & $0.25^{+0.30}_{-0.10}$ 
     & - & 7333 \\[3pt]
\hline\hline
\end{tabular*}
\end{table}

Three molecular states are found in the $J^P=1^+$ sector. The first, identified as a $B^* \olsi D$ bound state, is located 5, 2, 6, and 4 MeV below the $B^* \olsi D$ threshold in Models IA, IB, IIA, and IIB, respectively. This prediction is consistent with the LQCD results of Ref.~\cite{Alexandrou:2023cqg} as well as with previous phenomenological studies~\cite{Li:2012ss,Sakai:2017avl,Abreu:2022sra}. 
The second pole appears about $1\sim 2$ MeV below the $B\olsi D{}^*$ threshold for Model IA, IB, and IIA, with a mass splitting of roughly 95 MeV from the $B^*\olsi D$ threshold. In Model IIB, however, inclusion of the $D$-wave OPE potential shifts the pole to approximately $14$~MeV below the $B\olsi D{}^*$ threshold.\footnote{When the $D$-wave OPE potential is neglected, the pole position is $7287^{+1}_{-1}-0^{+1}_{-0}i$~MeV.}
This state possibly corresponds to the resonance identified in the $J^P=1^+$ $\bar b \bar c ud$ system, although it exhibits a significantly large decay width in the lattice study carried out in Ref.~\cite{Alexandrou:2023cqg}.  
The third pole lies tens of MeV below the $B^* \olsi D{}^*$ threshold.

\begin{table}[tb]
\caption{\label{Tab:pole_IIB} Same as Table~\ref{Tab:pole_IA}, but for Model IIB. }
\renewcommand{\arraystretch}{1.2}
\begin{tabular*}{\columnwidth}{@{\extracolsep\fill}lcccccc}
\hline\hline 
 $J^{PC}$ &  $H\olsi H$ & $M$ [MeV] & RS  & $g_{\text{low}}$ [GeV$^{-1/2}$] & $g_{\text{high}}$ [GeV$^{-1/2}$] & Thr. [MeV] \\[3pt]     
\hline
$0^{+}$  & $B\olsi D$ & $7146^{+0}_{-0}$
     & $(+)$ & $0.35^{+0.08}_{-0.10}$ & - & 7147 \\[3pt]
\hline
\multirow{2}{*}{$1^{+}$}  & \multirow{2}{*}{$B^*\olsi D$--$B\olsi D{}^*$} & $7188^{+1}_{-2}$ 
     & $(+,+)$ & $0.71_{-0.06}^{+0.07}$  
     & $-0.17_{-0.22}^{+0.28}$  
     & 7192 \\[3pt]
  & & $7274_{-2}^{+2}-7^{+1}_{-3}i$ 
     & $(-,+)$ & $0.03_{-0.03}^{+0.05}+0.05_{-0.06}^{+0.10}i$ 
     & $-1.1_{-0.0}^{+2.2}-0.33_{-0.11}^{+0.65}i$ 
     & 7288  \\[3pt]
\hline
$0^{+}$  & $B^*\olsi D{}^*$ & $7318^{+14}_{-23}$ 
     & $(+)$ &  $1.13^{+0.61}_{-0.69}$ 
     & - & 7333 \\[3pt]
\hline
$1^{+}$  & $B^*\olsi D{}^*$ & $7322_{-10}^{+~7}$ 
     & $(+)$ & $1.02^{+0.31}_{-0.35}$ 
     & - & 7333 \\[3pt]
\hline
$2^{+}$  & $B^*\olsi D{}^*$ & $7323^{+~6}_{-10

}$ & $(+)$ & $1.03^{+0.30}_{-0.25}$ 
     & - & 7333 \\[3pt]
\hline\hline
\end{tabular*}
\end{table}

The isoscalar $J^P=2^+$ molecular state lies below the $B^*\olsi D{}^*$ threshold by about $0\sim4$~MeV for Models IA, IB, and IIA. In Models IA and IB, this state is located on the second RS and is identified as a virtual state for all bootstrap samples within the $1\sigma$ region of the LECs. In Model IIA, it appears as a bound state in $63\%$ of the bootstrap samples, while $34\%$ of them yield a virtual state lying between the branch point of the left-hand cut (about 7332~MeV) and the threshold. In Model IIB, the state is bound by about 11~MeV, below the branch point of the left-hand cut. Notably, when the $D$-wave contribution is neglected, the binding energy decreases to 0.1~MeV.

In Ref.~\cite{Abolnikov:2026sik}, the possible pole positions have been predicted by reanalyzing the lattice spectra in Ref.~\cite{Alexandrou:2023cqg} under the two scenarios. In the first scenario, the poles have been determined using the contact potential $V=C_{0}+C_{2}(p^2+p^{\prime 2})$ for the single-channel $B^{(*)}\olsi D{}^{(*)}$ system, where $p$ and $p^\prime$ denote the initial and final three-momenta. The LECs, $C_{0}$ and $C_2$, are determined from the lowest four lattice spectra both in the $A_1^+$ and $T_1^+$ irreps using an exponential regulator with the cutoff $\Lambda=1.0$~GeV. In the second scenario, the contact potential has been extended to $V=C_0+C_2(p^2+p^{\prime 2})+C_4(p^4+p^{\prime 4})$ and the LECs have been determined by the nine and five lattice spectra in the $A_1^+$ and $T_1^+$ irreps with the cutoff $\Lambda=1.2$~GeV, respectively.\footnote{Here the potential $V$ corresponds to the $C_{0a}$ term in Eqs.~\eqref{Eq:contact_S0} and \eqref{Eq:contact_S1}, with $C_{0b}=0$.}
With the use of these two scenarios, the binding energies obtained for the $B\olsi D$, $B^*\olsi D$, and $B\olsi D{}^*$ systems in Ref.~\cite{Alexandrou:2023cqg} are generally consistent with our results in Tables~\ref{Tab:pole_IA}, \ref{Tab:pole_IB}, \ref{Tab:pole_IIA}, and \ref{Tab:pole_IIB},  once the uncertainties in the pole positions are taken into account. For the  $B^*\olsi D{}^*$ system, Ref.~\cite{Abolnikov:2026sik} sets $C_{0b}=0$ and predicts the degenerate $J=0,1,2$ states with identical binding energies.
Our results for the $1^+$ channel are compatible with those of Ref.~\cite{Abolnikov:2026sik}. For the $0^+$ and $2^+$ channels, some of our predictions are consistent with Ref.~\cite{Abolnikov:2026sik} within the quoted uncertainties, except for Models IA, IB, and IIA in the $0^+$ channel and Models IA and IB in the $2^+$ channel. In particular, Models IA and IB predict the $2^+$ state to be virtual rather than bound, whereas Models IIA and IIB give a bound state consistent with the result of Ref.~\cite{Abolnikov:2026sik}.

Furthermore, owing to the absence of nonlocal $B\olsi D{}^*$ interpolating operators in the lattice simulation of the $B^*\olsi D$--$B\olsi D{}^*$ system, the LEC $C_{0b}$ cannot be tightly constrained by the energy levels of the isoscalar $J^P=1^+$ $\bar b \bar c ud$ channel reported in Ref.~\cite{Alexandrou:2023cqg}. More precise lattice QCD results for the isoscalar $J^P=1^+$ $\bar b \bar c ud$ system, particularly including nonlocal $B\olsi D{}^*$ operators, will be essential for placing stronger constraints on the LECs and improving the precision of the predicted $B^{(*)}\olsi D{}^{(*)}$ molecular states.

\section{Summary}\label{Sec:Summary}

The observation of the doubly charmed tetraquark state in the $D^0D^0\pi^+$ invariant-mass distribution by the LHCb Collaboration~\cite{LHCb:2021auc,LHCb:2021vvq} has motivated us to investigate its bottom-charm partners with quark content $\bar b \bar c ud$. Guided by heavy quark spin and chiral symmetries, we construct the corresponding contact interactions, both with and without one-pion exchange (OPE) potentials, to study isoscalar $B^{(*)}\olsi D{}^{(*)}$ hadronic molecules with quantum numbers $J^P=0^+$, $1^+$, and $2^+$. The corresponding LECs are determined by reanalyzing the LQCD energy levels reported in Ref.~\cite{Alexandrou:2023cqg} for the isoscalar $\bar b \bar c ud$ channels with $J^P=0^+$ and $1^+$. The spectrum of the resulting isoscalar doubly-heavy tetraquark states is then predicted within the hadronic molecular framework.

To reanalyze the LQCD data, we consider four models. Models IA and IB include only short-range contact interactions, with cutoff values of $\Lambda=0.8$ and $1.2$ GeV, respectively, while Models IIA and IIB additionally incorporate the long-range OPE potential with the corresponding cutoff values. For each of these models, we search for poles in the $B^{(*)}\olsi D{}^{(*)}$ systems. We predict shallow bound states in the $J^P=0^+$ $B\olsi D$ and $J^P=1^+$ $B^*\olsi D$ channels, with binding energies of only a few MeV, in agreement with the  LQCD results of Ref.~\cite{Alexandrou:2023cqg}. 

Although the masses of the $J^P=0^+$ and $1^+$ molecular states near the $B^*\olsi D{}^*$ and $B\olsi D{}^*$ thresholds are compatible with those of the corresponding resonances obtained in the lattice QCD analysis of Ref.~\cite{Alexandrou:2023cqg}, the large decay widths of these resonances indicate that further lattice QCD studies of the isoscalar $\bar b \bar c ud$ system are required. In particular, the inclusion of nonlocal $B^*\olsi D{}^{*}$ and $B\olsi D{}^*$ interpolating operators in future lattice simulations would allow a more reliable determination of the pole positions and provide stronger constraints on the $B^{(*)}\olsi D{}^{(*)}$ hadronic molecular spectrum.

\begin{acknowledgments}
We are grateful to Ming-Zhu~Liu, Jia-Jun~Wu, Jia-Jun~Xie, and Kang~Yu for useful discussions. This work is part of the Grants {\small PID2023-147458NB-C21} and {\small CEX2023-001292-S} funded by {\small MICIU/AEI/10.13039/501100011033} and by ERDF/EU, as well as of the Grants {\small CIPROM/2023/59} and HADRON-2030 funded by Generalitat Valenciana {\small 10.13039/501100003359} and the European Unison’s Horizon Europe research and innovation programme under agreement No. 101292522, respectively. This work is also supported in part by the National Natural Science Foundation of China (NSFC) under Grants No. 12125507, No. 12361141819, No. 12447101, No. 12675095, and No. 12547111; by the Chinese Academy of Sciences (CAS) under Grant No. YSBR-101; and by the National Key R\&D Program of China under Grant No. 2023YFA1606703. M.\,A. acknowledges financial support through GenT program by Generalitat Valencia (GVA) Grant No.\,CIDEGENT/2020/002, Ramón y Cajal program by MICINN Grant No.\,RYC2022-038524-I, and Atracción de Talento program by CSIC Grant No.\, PIE 20245AT019. P.-P. Shi is grateful to the Institute of Theoretical Physics, CAS, and the University of Electronic Science and Technology of China for the hospitality during his visit. V. Montesinos thanks the University of Electronic Science and Technology of China for the hospitality.
\end{acknowledgments}

\appendix

\section{\boldmath Partial-wave-projected potentials for the $B^*\olsi D{}^*$ system}\label{sec:partial_wave}
\label{sec:AppendixA}

The OPE potentials of the  $B^*\olsi D{}^*$ system can be decomposed into partial-wave components according to the quantum numbers $J^P$. For the $J^P=0^+$ sector, we consider the coupled $S$- and $D$-wave channels, $\left\{B^*\olsi D{}^*({\cal S}),\,B^*\olsi D{}^*({\cal D})\right\}$.
The corresponding OPE potential is given by
\begin{align}
V_{\OPE}^0(\boldsymbol{q},B^*\olsi D{}^*)&=\frac{1}{8}\frac{g^2}{F_{\pi}^2}\begin{pmatrix}
& 2V_{\cal SS}(p,k) &-V_{\cal SD}(p,k)\\
& -V_{\cal DS}(p,k) &V^0_{\cal DD}(p,k)/2\\
\end{pmatrix}.
\end{align}
For the $1^+$ sector, the relevant partial-wave basis is
\begin{align}
\left\{B^*\olsi D{}^*({\cal S}),\, B^*\olsi D{}^*({\cal D}^1),\, B^*\olsi D{}^*({\cal D}^2) \right\},
\end{align}
where the superscript $S_{\text{tot}}$ in ${\cal D}^{S_{\text{tot}}}$ denotes the total spin of the $B^*\olsi D{}^*$ system. In this basis, the OPE potential reads
\begin{align}
V_{\OPE}^1(\boldsymbol{q},B^*\olsi D{}^*)&=\frac{1}{8}\frac{g^2}{F_{\pi}^2}\begin{pmatrix}
& V_{\cal SS}(p,k) &V_{\cal SD}(p,k) & 0 \\
& V_{\cal DS}(p,k) &V_{\cal DD}(p,k) & 0\\
& 0 & 0  & 3V^1_{\cal DD}(p,k)/2 \\
\end{pmatrix}.
\end{align}
In the basis $\left\{B^*\olsi D{}^*({\cal S}), \, B^*\olsi D{}^*({\cal D}^0),\, B^*\olsi D{}^*({\cal D}^1),\, B^*\olsi D{}^*({\cal D}^2) \right\}$,
the OPE potential in the $2^+$ sector is
\begin{align}
V_{\OPE}^2(\boldsymbol{q},B^*\olsi D{}^*)&=\frac{1}{8}\frac{g^2}{F_{\pi}^2}\begin{pmatrix}
& -V_{\cal SS}(p,k) &-V_{\cal SD}(p,k)/\sqrt{5} & 0 & \sqrt{7/5}V_{\cal SD}(p,k) \\
& -V_{\cal DS}(p,k)/\sqrt{5} & V^2_{\cal DD}(p,k) & 0 & -V^3_{\cal DD}(p,k)/\sqrt{7}\\
& 0 & 0  & -3V^1_{\cal DD}(p,k)/2  & 0 \\
& \sqrt{7/5}V_{\cal DS}(p,k) & -V^3_{\cal DD}(p,k)/\sqrt{7}  & 0 & V_{\cal DD}^4(p,k)
\end{pmatrix}\,.
\end{align}
Here the individual partial-wave-projected components $V_{\cal SS}(p,k)$, $V_{\cal SD}(p,k)$, $V_{\cal SS}(p,k)$, and $V_{\cal DD}(p,k)$ are defined in Eq.~\eqref{Eq:partial_wave_component}. The additional $D$-wave components are given by
\begin{align}
V^0_{\cal DD}(p,k)&=\hphantom{+}\int_{-1}^1 \text{d}z\, D_{\pi}(E)\left[{(k^2+p^2)(3z^2-1)-4pkz}\right],
\nonumber\\
V^1_{\cal DD}(p,k)&=\hphantom{+}\int_{-1}^1 \text{d}z\, D_{\pi}(E)\left[{pkz(z^2-1)}\right],\nonumber\\
V^2_{\cal DD}(p,k)&=\hphantom{+}\int_{-1}^1 \text{d}z\, D_{\pi}(E)\left[{(k^2+p^2)(3z^2-1)+2pkz(1-3z^2)}\right],\nonumber\\
V^3_{\cal DD}(p,k)&=\hphantom{+}\int_{-1}^1 \text{d}z\, D_{\pi}(E)\left[{(k^2+p^2)(3z^2-1)-pkz(1+3z^2)}\right],\nonumber\\
V^4_{\cal DD}(p,k)&=-\int_{-1}^1 \text{d}z\, D_{\pi}(E)\left[{10(k^2+p^2)(3z^2-1)+pkz(11-51z^2)}\right]/14.
\end{align}

\section{\boldmath Statistical correlation matrices for the coupled $B^*\olsi D$ and $B\olsi D^*$ channels}\label{sec:AppendixB}

In this appendix, we present the correlation matrices for the four fitting scenarios considered in this work. The corresponding values of the LECs are listed in Table~\ref{Tab:LECs}. The resulting symmetric correlation matrices are given by
\begin{equation}
\begin{matrix}
 & C_{0a}  & C_{0b}   \\
C_{0a} \ldelim[{2}{0.1cm} & 1.00  & 0.62  & \rdelim]{2}{0.1cm}\\
C_{0b} & 0.62    & 1.00  \\
\end{matrix}\qquad\qquad
\begin{matrix}
 & C_{0a}  & C_{0b}   \\
C_{0a} \ldelim[{2}{0.1cm} & 1.00  & 0.64  & \rdelim]{2}{0.1cm}\\
C_{0b} & 0.64    & 1.00  \\
\end{matrix}
\end{equation}
for Model IA and IB, and
\begin{equation}
\begin{matrix}
 & C_{0a}  & C_{0b}   \\
C_{0a} \ldelim[{2}{0.1cm} & 1.00  & 0.59  & \rdelim]{2}{0.1cm}\\
C_{0b} & 0.59    & 1.00  \\
\end{matrix}\qquad\qquad
\begin{matrix}
 & C_{0a}  & C_{0b}   \\
C_{0a} \ldelim[{2}{0.1cm} & 1.00  & 0.21  & \rdelim]{2}{0.1cm}\\
C_{0b} & 0.21    & 1.00  \\
\end{matrix}
\end{equation}
for Model IIA and IIB.

\bibliography{ref.bib}
\end{document}